\documentclass[12pt,a4]{article}
\usepackage{graphicx} 
\usepackage{amsfonts}
\usepackage{amssymb}
\usepackage{amsmath}
\usepackage{siunitx}
\usepackage{tikz}
\usepackage{pgfplots}
\usetikzlibrary{intersections, calc, decorations.markings}
\usepackage[font=small,labelfont=bf]{caption}
\def\be{\begin{eqnarray}}
\def\ee{\end{eqnarray}}

\usetikzlibrary{arrows.meta, decorations.markings, calc, angles, quotes}
\usepackage{mathrsfs}
\title{Minimal Length Effects on Keplerian Scattering and Gravitational Lensing}
\author{
Mykola Samar, Mariia Seniak\\[4pt]
\small Professor Ivan Vakarchuk Department for Theoretical Physics,\\
\small Ivan Franko National University of Lviv,\\
\small 12 Drahomanov St., Lviv, UA 79005, Ukraine
}

\begin{document}

\maketitle

\begin{abstract}
We study the impact of a minimal length, implied by generalized uncertainty principles and quantum gravity models, on unbounded (scattering) trajectories in the Kepler problem. The analysis is based on the precession of the Hamilton vector, which serves as a sensitive probe of orbital perturbations. Within the framework of the deformed Heisenberg algebra, we derive the correction to the trajectory arising from minimal length effects. It is shown that these quantum-gravitational corrections lead to a reduction in the scattering angle. In particular, for massless particles such as photons, the quantization of space results in a weakening of the gravitational lensing effect. Using available experimental data from the observation of the Einstein ring, we estimate the deformation parameter and the corresponding minimal length for the electron and Mercury. These findings highlight potential observational signatures of minimal length scenarios in high-energy astrophysics and gravitational optics.
\end{abstract}

\section{Introduction}
Recently, considerable attention has been devoted to quantum mechanical problems involving generalized (modified) commutation relations. This interest has been largely driven by developments in quantum gravity and string theory, which suggest the existence of a finite lower bound on spatial resolution — the so-called minimal length~\cite{GrossMende,Maggiore,Witten}. Kempf and collaborators demonstrated that a minimal length naturally arises from a deformation of the Heisenberg algebra~\cite{Kempf1994,KempfManganoMann,HinrichsenKempf,Kempf1997}. It is worth noting that such a deformed algebra, leading to quantized space-time, was originally introduced by Snyder in the relativistic context~\cite{Snyder}.

In the $D$-dimensional case, the deformed algebra proposed by Kempf takes the form
\begin{align}
[X_i, P_j] &= i\hbar \left( \delta_{ij}(1 + \beta P^2) + \beta' P_i P_j \right), \nonumber \\
[P_i, P_j] &= 0, \label{eq:kempf-algebra} \\
[X_i, X_j] &= i\hbar \frac{(2\beta - \beta') + (2\beta + \beta')\beta P^2}{1 + \beta P^2} (P_i X_j - P_j X_i), \nonumber
\end{align}
where $\beta$ and $\beta'$ are non-negative deformation parameters: $\beta, \beta' \geq 0$. As seen from Eq.~\eqref{eq:kempf-algebra}, the position operators do not commute, resulting in a noncommutative geometry. 
The corresponding generalized uncertainty relation implies the existence of a minimal length given by \begin{equation}\ell_{min}=\label{eq:minimal length}
\hbar \sqrt{D\beta + \beta'}.
\end{equation}  
Interestingly, in the special case $\beta' = 2\beta$, the commutator of position operators vanishes to first order in the deformation parameters, i.e., $[X_i, X_j] = 0$. 

In the classical limit $\hbar \to 0$, the quantum mechanical commutator between operators is replaced by the Poisson bracket of the corresponding classical variables:
\begin{equation}
\frac{1}{i\hbar} [X, P] \ \longrightarrow \ \{X, P\}. 
\end{equation}
For the deformed algebra considered here, the Poisson bracket takes the form
\begin{align}
\{X_i, P_j\} &= \delta_{ij}(1 + \beta P^2) + \beta' P_i P_j, \nonumber \\
\{P_i, P_j\} &= 0, \label{Poisson} \\
\{X_i, X_j\} &= \frac{(2\beta - \beta') + (2\beta + \beta')\beta P^2}{1 + \beta P^2} (P_i X_j - P_j X_i). \nonumber
\end{align}
which define the deformed phase-space structure associated with the generalized uncertainty principle.

The framework of the minimal length hypothesis has been extensively applied to a wide range of quantum mechanical systems. Among them are the harmonic oscillator~\cite{KempfManganoMann, Tkachuk2003, Tkachuk2004, Chang, Dadic}, the Dirac oscillator~\cite{Tkachuk2005, Tkachuk2006}, the hydrogen atom~\cite{Brau, Benczik, StetskoTkachuk, Stetsko,  SamarTkachuk, Samar}, the gravitational quantum well~\cite{Brau2006, Pedram2011}, a particle in a delta potential~\cite{Samar1, Ferkous}, one-dimensional Coulomb-like systems~\cite{Samar1, Fityo, Samar2}, particles in singular inverse-square potentials~\cite{Bouaziz2007, Bouaziz2008}, the Casimir effect~\cite{Frassino}, and scattering processes~\cite{Stetsko2007}.

The effects of space quantization have also been investigated at the classical level for various physical systems, including Keplerian motion, statistical systems, and composite bodies~\cite{Benczik2002, Silagadze,   Fityo2008, Frydryszak, Buisseret,   Tkachuk2010,Tkachuk2012, Samar2020, Kovach2022}.

The deformed Heisenberg algebra that incorporates a minimal length provides a convenient phenomenological framework for describing quantized space. However, this approach gives rise to several fundamental issues, one of which is the violation of the weak equivalence principle~\cite{Tkachuk2012}. This difficulty emerges if the deformation parameter is assumed to be universal, i.e., identical for both elementary particles and macroscopic bodies.

To overcome this problem, it was proposed in~\cite{Tkachuk2010, Tkachuk2012, Samar2020} to relate the deformation parameter to the particle mass according to
\begin{equation}\label{g_cond}
\beta = \frac{\gamma}{m^2},
\end{equation}
where $\gamma$ is a universal constant. This assumption restores the weak equivalence principle in deformed space with minimal length and ensures the additivity of kinetic energy for composite systems, as well as its independence from internal structure.

Despite the considerable number of studies devoted to the minimal length hypothesis, its experimental verification remains an open problem. The Kepler problem is particularly suitable for this purpose due to the high precision of both theoretical predictions and observations. Unbounded (scattering) trajectories provide a sensitive probe of small deviations from standard dynamics, as they are strongly affected by perturbations of the interaction and phase space structure. A convenient tool for such analysis is the Hamilton vector, whose precession leads to corrections in the scattering angle \cite{Chashchina}.

An important application of scattering processes in gravitational fields is gravitational lensing. The deflection of light by massive bodies provides one of the most precise tests of gravitational theories and has been confirmed by numerous observations. In particular, strong lensing phenomena, such as Einstein rings, offer accurate measurements of the deflection angle.

The presence of a minimal length is expected to affect the propagation of light in gravitational fields. This leads to corrections to the deflection angle and, consequently, to observable deviations in gravitational lensing patterns. Therefore, gravitational lensing serves as a promising tool for probing quantum-gravitational effects associated with the existence of a minimal length. In this context, it is of particular interest to analyze how the deformation of the phase space structure influences the bending of light and to estimate the corresponding constraints on the deformation parameter from observational data.

The paper is organized as follows. In Sec. 2, we formulate the problem and derive the correction to the scattering angle. In Sec. 3, we discuss gravitational lensing in the presence of a minimal length, observational implications, and provide estimates of the minimal length. Conclusions are given in Sec. 4.

\section{Perturbation of the scattering angle}
We study the Kepler problem
\begin{equation}
H = \frac{P^2}{2m} - \frac{\alpha}{R},
\label{eq:Hamiltonian}
\end{equation}
where the generalised radial coordinate is defined as
\(
R = \sqrt{\sum_{i=1}^3 X_i^2}
\) and the position $X_i$ and momentum $P_i$ obey the deformed  Poisson bracket given by Eq.(\ref{Poisson}). We focus on the scattering problem rather than on bound orbits, which were considered earlier in \cite{Benczik, Silagadze}.
The strength of the gravitational interaction is characterized by
\be\label{alpha}
\alpha = GmM,
\ee
where \( G \) is the gravitational constant, \( m \) is the mass of the incoming particle, and \( M \) is the mass of the attractive center. 
We adopt the following representation for the position and momentum  which satisfies the deformed Poisson brackets given in equation (\ref{Poisson}) up to first order in the deformation parameters $\beta$ and $\beta'$:
\begin{equation} \label{eq:representation}
\begin{cases}
X_i = x_i\left(1 + \frac{2\beta - \beta'}{2} p^2\right), \\
P_i = p_i\left(1 + \frac{\beta'}{2} p^2\right),
\end{cases}
\end{equation}
where the squared momentum is defined as $p^2 = \sum_{k=1}^3 p_k^2$, and the variables $x_i$, $p_j$ satisfy the standard canonical Poisson brackets:
\be \{x_i, p_j\} = \delta_{ij}.
\ee
We now express the Hamiltonian from equation~(\ref{eq:Hamiltonian}) using the representation given in equation~(\ref{eq:representation}), keeping only terms linear in the deformation parameters $\beta$ and $\beta'$.  This leads to the Hamiltonian in the form:
\begin{equation}
H = \frac{p^2}{2m} + \frac{\beta'}{2m} p^4 - \frac{\alpha}{r} \left( 1 - \frac{2\beta - \beta'}{2} p^2\right). \label{eq:H_expanded}
\end{equation}
The Hamiltonian can be rewritten as
\begin{equation}
H = H_0+\Delta H_\beta,
\end{equation}
where  the unperturbed Hamiltonian and perturbation correspondingly are 
\begin{align}
H_0 &= \frac{p^2}{2m} - \frac{\alpha}{r}, 
\label{eq:H0} \\
\Delta H_\beta &= \frac{\beta'}{2m} p^4 
+ \frac{2\beta - \beta'}{2} \frac{\alpha p^2}{r}.
\label{eq:DeltaHbeta}
\end{align}

Perturbations in the central potential lead to a slow rotation of the orbit, resulting in the precession of the perihelion. A particularly elegant method for analyzing this effect involves the use of vector constants of motion, such as the Hamilton vector~\cite{ Silagadze,Chashchina}, whose precession rate coincides with that of the perihelion. The Hamilton vector is defined as~\cite{Chashchina}
\begin{equation}
\vec{u} = \frac{\vec{p}}{m} 
- \alpha \frac{ \vec{L} \times \vec{r} }{L^2 r},
\label{eq:HamiltonVector}
\end{equation}
and is conserved in the absence of perturbations:
\be
\dot{\vec{u}} = \{\vec{u}, H_0\} = 0.
\ee
Here, $L = m r^2 \dot{\varphi}$ is the orbital angular momentum.

In the presence of the perturbation, the Hamilton vector is no longer conserved:
\be
\dot{\vec{u}} = \{\vec{u}, H\} \neq 0.
\ee
Its time derivative is given by
\be
\dot{\vec{u}} = \{\vec{u}, H\} = \{\vec{u}, \Delta H_\beta\}.
\ee
By substituting expressions from equations (\ref{eq:DeltaHbeta}) and ~(\ref{eq:HamiltonVector}), we find
\be
\dot{\vec{u}} = \left(\frac{(2\beta+3\beta')\alpha p^2}{2m r^3}+\frac{(2\beta-\beta')\alpha^2}{r^4}\right) \vec{r}. 
\ee
The angular velocity of precession of the Hamilton vector is defined as
\be
\vec{\omega} = \frac{\vec{u} \times \dot{\vec{u}}}{|\vec{u}|^2}.
\ee
Using the vector identity $(\vec{L} \times \vec{r}) \times \vec{r} = -r^2 \vec{L}$ and the relation $|\vec{u}| = \frac{\alpha e}{L}$, where $
e $ denotes the eccentricity, we obtain the expression
\be
\vec{\omega} = \left( \frac{(2\beta + 3\beta') p^2}{2m r^2} + \frac{(2\beta - \beta')\alpha}{r^3} \right) \left(1 - \frac{ \mathrm{p} }{r} \right) \frac{\vec{L}}{e^2},
\ee
 where the quantity
\be
 \mathrm{p} =\frac{L^2}{m\alpha}\ee
 denotes the semi-latus rectum of the unperturbed orbit.
 
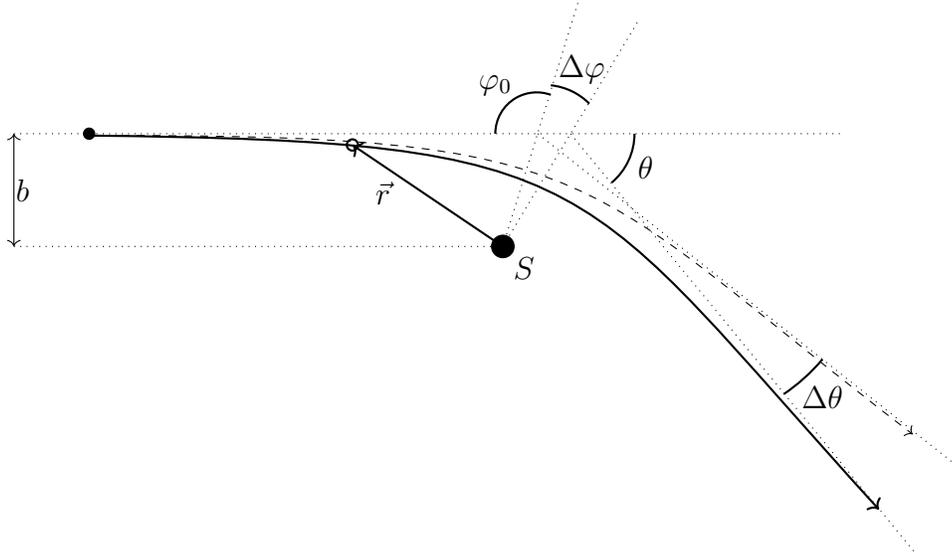
\begin{figure}[h!]
\centering
\begin{tikzpicture}[scale=5]


\filldraw (0,0) circle (0.03) node[below right] {$S$};

\draw[thick, ->] (-1.1,0.295) .. controls (0.28,0.28) and (0.19,0.19) .. (1,-0.7);

\draw[dashed,->] (-1.1,0.295) .. controls (0.3,0.3) and (0.19,0.19) .. (1.09,-0.5);

\draw[<->] (-1.3,0) -- (-1.3,0.3);
\node[left] at (-1.23,0.15) {$b$};


\draw[dotted] (-1.3,0.3) -- (0.9,0.3);
\filldraw (-1.1,0.3) circle (0.015) node[left] {};

\draw[dotted] (-1.3,0) -- (0,0);

\draw[dotted] (0.18,0.3) -- (1.1,-0.82);

\draw[dotted] (0.093,0.3) -- (1.21,-0.586);

\draw[-, thick] (0.35,0.3) arc[start angle=0, end angle=-50, radius=0.17];
\node at (0.38,0.21) {$\theta$};

\draw[-, thick] (0.85,-0.3) arc[start angle=-40, end angle=-56, radius=0.5];
\node at (0.85,-0.4) {$\Delta\theta$};

\draw[-, thick] (0.23,0.38) arc[start angle=45, end angle=84, radius=0.17];
\node at (0.21,0.47) {$\Delta \varphi$};

\draw[dotted] (0,0) -- (0.36,0.6);
\draw[dotted] (0,0) -- (0.20,0.65);

\coordinate (rvec) at (-0.4,0.27);
\coordinate (Lvec) at (-0.35,0.4);
\draw[->, thick] (0,0) -- (rvec);
\node[below left] at (-0.27,0.2) {$\vec{r}$};
\draw[thick] (-0.4,0.27) circle (0.015) node[left] {};

\draw[-, thick] (-0.02,0.3) arc[start angle=180, end angle=70, radius=0.11];
\node at (-0.02,0.43) {$\varphi_0$};

\end{tikzpicture}
\caption{Scattering trajectory near a central force center \( S \). The impact parameter is denoted by \( b \), and the scattering angle is \( \theta \). The change in scattering angle due to perturbation is \( \Delta\theta \), while \( \Delta\varphi \) represents the deviation of the angle between the major axis of the hyperbola and the asymptote from its unperturbed value \( \varphi_0 \).}
\label{fig:scattering}
\end{figure}

Consequently, the precession of the Hamilton vector—equivalently, the perihelion shift—accumulated during the motion from the point of closest approach to infinity  reads
\begin{equation}\label{integral of an angle of rotation of the perihelion}
\Delta\varphi = \int_0^\infty \omega \, dt = \int_0^{\varphi_{0}} \frac{\omega}{\dot{\varphi}} \, d\varphi,  
\end{equation}
where $\varphi_0$ denotes the angle between the direction of the perihelion (the axis of symmetry of the scattering trajectory) and the asymptote of the unperturbed trajectory ({\it see Fig.\ref{fig:scattering}}).
Here the angular velocity of motion in the linear approximation on deformation parameters $\beta$ and $\beta'$ is equal to
\be\dot{\varphi}=\frac{L}{mr^2}.\ee
From this, we obtain
\be
\frac{\omega}{\dot{\varphi}}=\left( \frac{(2\beta + 3\beta') p^2}{2e^2} + \frac{(2\beta - \beta')\alpha m}{e^2r} \right) \left(1 - \frac{ \mathrm{p} }{r} \right). 
\ee

To evaluate the integral in equation (\ref{integral of an angle of rotation of the perihelion}) to first order in the perturbation, we may utilize expressions that hold for the unperturbed orbital motion:
\be 
r = \frac{\mathrm{p}}{1 + e \cos \varphi}, \quad \text{and} \quad E=\frac{p^2}{2m} - \frac{\alpha}{r} = \frac{\alpha(e^2-1)}{2\mathrm{p}}.
\ee
We consider the hyperbolic scattering orbits with $e>1$, the unperturbed  value of angle $\varphi_{0}$ is equal to
\be \label{phi0} \varphi_{0}=\pi-\arccos\left(\frac{1}{e}\right).\ee
Using these relations, we can proceed to compute the integral
\be 
\Delta\varphi = -\frac{\alpha m}{e\mathrm{p}} \int_0^{\varphi_{0}} \left [ \frac{(6\beta+\beta')+(2\beta+3\beta')e^2}{2}+(4\beta+2\beta')e \cos \varphi\right]\cos \varphi d\varphi
.\ee

After the integration, we finally obtain
\be \label{dp} \Delta\varphi=
- \frac{ \alpha m }{ b } 
\left(\frac{2\beta+3\beta'}{2}+\frac{2\beta+\beta'}{\sqrt{e^2-1}}\left(\pi-
 \arccos\left( \frac{1}{e} \right) \right)
+ \frac{2\beta-\beta'}{2e^2}
\right).
\ee

Here, we have introduced the impact parameter \(b\), which is related to the semi-latus rectum \(\mathrm{p}\) of the hyperbolic orbit through the following relation:
\begin{equation}
b = \frac{\mathrm{p}}{\sqrt{e^2-1}} .
\end{equation}
As shown in Fig.~1, the scattering angle including the effects of deformation is given by
\begin{equation}
\theta = \theta_0 + \Delta\theta.
\end{equation}
Here, $\theta_0$ is the scattering angle in the undeformed Kepler problem
\begin{equation}
\theta_0 = 2\varphi_{0} - \pi = \pi - 2\arccos\left(\frac{1}{e}\right),
\end{equation}
while $\Delta\theta$ denotes the correction due to minimal-length effects
\begin{equation}
\label{DTheta}
\Delta\theta = 2\Delta\varphi = - \frac{G M m^2}{b} \left( 
2\beta + 3\beta'
+ \frac{4\beta + 2\beta'}{\sqrt{e^2 - 1}} \left(\pi - \arccos\left(\frac{1}{e}\right)\right)
+ \frac{2\beta - \beta'}{e^2}
\right),
\end{equation}
where $\alpha$ from Eq.~(\ref{alpha}) has been substituted.

It is important to emphasize that the obtained result can also be consistently extended to the case of parabolic trajectories ($e = 1$). In this limit, one obtains
\begin{equation}
\Delta\theta = - \frac{GMm^2}{\mathrm{p}}\,(4\beta+2\beta')\,\pi.
\end{equation}
Thus, we conclude that the scattering angle for unbounded (scattering) trajectories is reduced due to minimal-length effects.

However, we observe that the scattering angle explicitly depends on the mass of the scattered particle, which constitutes a violation of the \textit{weak equivalence principle}—the assertion that the motion of all bodies in a gravitational field is independent of their mass. Nevertheless, this issue can be readily resolved by assuming that the deformation parameters (and, consequently, the minimal length) are themselves mass-dependent:
\begin{equation}\label{beta}
\beta = \frac{\gamma}{m^2}, \qquad \beta' = \frac{\gamma'}{m^2},
\end{equation}
where \(\gamma\) and \(\gamma'\) are fundamental (mass-independent) constants.
After substituting Eq.~(\ref{beta}) into Eq.~(\ref{DTheta}), we obtain the minimal-length correction to the scattering angle
\begin{equation}\label{delta theta}
\Delta\theta =  - \frac{G M}{b} 
\left(
2 \gamma + 3 \gamma'
+ \frac{4 \gamma + 2 \gamma'}{\sqrt{e^2 - 1}} \left( \pi - \arccos \left( \frac{1}{e} \right) \right)
+ \frac{2 \gamma - \gamma'}{e^2}
\right),
\end{equation}
which is independent of the mass of the incoming particle. 
We stress that the eccentricity of the unperturbed orbit does not depend on the mass of the particle
\be \label{eccenticity}
e = \sqrt{1 + \frac{2 E L^2}{m \alpha^2}} =
\sqrt{1 + \frac{2 \cdot \frac{m v_\infty^2}{2} \cdot (m v_\infty b)^2}{m (G M m)^2}} =
\sqrt{1 + \frac{v_\infty^4 b^2}{G^2 M^2}}.
\ee
Thus, the {weak equivalence principle} is restored.

The concept of a {mass-dependent minimal length} was previously explored in~\cite{Tkachuk2010, Tkachuk2012} as a way to resolve several fundamental issues introduced by the notion of minimal length, namely: the violation of the weak equivalence principle, the composition dependence of kinetic energy, the discrepancy in minimal length estimations from planetary motion, and the mass-dependent Galilean and Lorentz transformations. It is also worth noting that a similar approach has been successfully applied in the context of relativistic versions of deformed algebra~\cite{Samar2020, Kovach2022}, as well as in the framework of noncommutative geometry~\cite{Gnatenko2013, Gnatenko2017}. This further supports the idea and suggests that it may be rooted in a more fundamental physical principle.

\section{Gravitational lensing in the presence of a minimal length}
The phenomenon of gravitational lensing arises when a massive object bends the path of light coming from a distant source, effectively acting as a lens and amplifying the observed brightness. 
The deflection of light by gravity was predicted by Einstein in 1915 as part of his general theory of relativity~\cite{ref11}. Later, in 1936, he extended this idea by suggesting that a massive body could act as a lens,  enhancing the brightness of a background star by bending its light~\cite{ref12}.

\begin{figure}[ht]
\centering
\begin{tikzpicture}[scale=2.0]

\filldraw (2,0) circle (0.03) node[below right] {O};

\filldraw (0,0) circle (0.05) node[below right, yshift=2pt] {L};

\filldraw (-3,0) circle (0.03) node[below right] {S};

\draw[dashed, thick] (-3,0) ellipse [x radius=0.35, y radius=1.5];

\draw[dotted] (0,-1) -- (0,1);
\draw[dotted] (-3,0) -- (0,0) node[midway, fill=white] {$d_{LS}$};
\draw[dotted] (0,0) -- (2,0) node[midway, fill=white] {$d_{L}$};
\draw[dotted] (-3,-1.5) -- (2,-1.5) node[midway, fill=white] {$d_{S}$};
\draw[dotted] (2,0) -- (2,-1.5);
\draw[dotted] (-3,0) -- (1,0.8);
\draw[dotted] (-3,0) -- (1,-0.8);
\draw[dotted] (-3,1.5) -- (2,0);
\draw[dotted] (-3,-1.5) -- (2,0);

\draw[thin] (0,0) -- (0,0.6) node[midway, fill=white] {$b$};
\draw[thick] (-3,0) -- (0,0.6);
\draw[thick] (-3,0) -- (0,-0.6);
\draw[thick] (0,0.6) -- (2,0);
\draw[thick] (0,-0.6) -- (2,0);

\draw[thin] (0,0.6) ++(-18:0.45) arc (-18:12:0.45);
\node at (0.6,0.6) {\small $\theta$};

\draw[thin] (2,0) ++(163:0.35) arc (163:180:0.35);
\draw[thin] (2,0) ++(163:0.38) arc (163:180:0.38);
\node at (1.4,0.08) {\small $\alpha_E$};

\draw[thin] (-3,0) ++(0:0.8) arc (0:11:0.8);
\draw[thin] (-3,0) ++(0:0.82) arc (0:11:0.82);
\draw[thin] (-3,0) ++(0:0.84) arc (0:11:0.84);
\node at (-2,0.1) {\small $\phi$};

\end{tikzpicture}

\caption{Formation of an Einstein ring when the source $S$, the lens $L$, and the observer $O$ are perfectly aligned. The dashed ellipse represents the Einstein ring. The angle $\alpha_E$ denotes the Einstein angle. The angle $\theta$ is the deflection angle caused by the lens. The angle $\phi$ is measured between the original (unlensed) direction of the light and the line connecting the source to the lens.}
\label{fig:einstein_ring}
\end{figure}
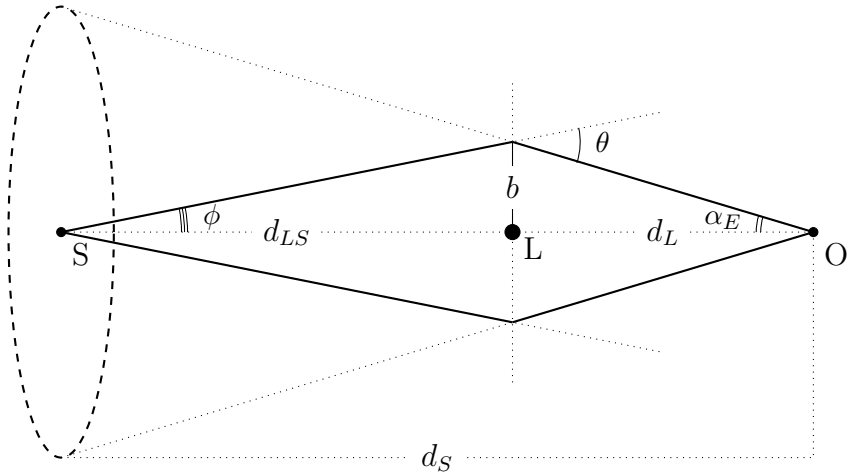

A simple example of gravitational lensing is the Einstein ring, which occurs when the source, lens, and observer are perfectly aligned. In this case, the deflected light forms a ring-like image centered around the lensing mass \cite{ref14}. Fig.~\ref{fig:einstein_ring} illustrates the formation of an Einstein ring. Let $S$ denote the source, $L$ the lens, and $O$ the observer. The angle $\alpha_E$, known as Einstein angle, represents the apparent angular radius of the Einstein ring as seen by the observer, while $\varphi$ is the angle between the unperturbed light ray and the straight line connecting the source to the lens. By the exterior angle theorem, the {angle of deviation} $\theta$, i.e., the angle between the original and deflected light rays, is given by
\begin{equation}
    \theta = \alpha_E + \varphi.
\end{equation}
As it can be seen from Fig.~\ref{fig:einstein_ring} \[
\alpha_E \approx \tan \alpha_E = \frac{b}{d_L}, \qquad
\phi \approx \tan \phi = \frac{b}{d_{LS}}.
\]
Here, $d_L$, $d_S$, and $d_{LS} = d_S - d_L$ denote the observer--lens, 
observer--source, and lens--source distances, respectively.
From this we obtain
\begin{equation} \label{theta}
\theta = \alpha_E \left( 1 + \frac{d_L}{d_{LS}} \right).
\end{equation}


If we wish to include the effect of a minimal length on the formation of Einstein rings, 
we note that for a photon one can formally set $v_\infty = c$ in the expression for the eccentricity (\ref{eccenticity}) of the unperturbed orbit. In this case, the eccentricity becomes very large,
corresponding to a trajectory that is nearly straight.
In the limit $e \gg 1$, the scattering angle in ~(\ref{delta theta}) simplifies to
\begin{equation}\label{DT_light}
\Delta \theta \approx - \frac{G M}{b} \left(2 \gamma + 3 \gamma'\right).
\end{equation}
We want to stress that, although this is a {naive extrapolation} of the classical formula to massless particles, it provides a reasonable order-of-magnitude estimate. In the undeformed case, the classical formula slightly underestimates the bending of light, differing only by a factor of two from the general-relativistic result:
\begin{equation}\label{GR_light}
\theta_0^{\rm GR} = 2 \, \theta_0^{\rm class} = \frac{4 G M}{c^2 b}.
\end{equation}
Extending this approach to the deformed case, we expect to obtain a scattering angle that remains close to the true value. Of course, the exact deflection of light in a gravitational field requires the framework of general relativity, which fully accounts for the curvature of spacetime and involves a significantly more complex analysis. Thus, despite its simplicity, our approach provides a reasonably accurate estimate for practical purposes and can be used for order-of-magnitude calculations.

We now estimate the scattering angle of light passing near a gravitational lens. Here, the lens is a star whose gravitational field deflects nearby light rays. The parameters of the system are taken from experimental data~\cite{ref13}, obtained from observations of the Einstein ring associated with Stein~2051, and are given by
\begin{equation}
d_L = 5.52 \pm 0.01~\mathrm{pc}, \quad 
d_S = 2 \times 10^3~\mathrm{pc}, \quad  \alpha_E = 31.53 \pm 1.20~\mathrm{mas}.
\end{equation}
According to (\ref{theta}) the scattering angle  is
\begin{equation}\label{theta_exp}
\theta = 31.62 \pm 1.20~\mathrm{mas}.
\end{equation}
Note that the influence of uncertainties in $d_L$ and $d_{LS}$ on the angle $\theta$ is negligible due to the smallness of the ratio $d_L/d_{LS} \ll 1$; therefore, the error in $\theta$ is practically determined solely by the measurement uncertainty of $\alpha$.

Assuming that the absolute value of the minimal length correction to the deflection angle from (\Ref{DT_light}) is smaller than the measurement uncertainty \be|\Delta\theta|\leq \Delta\theta^{exp}=1.20~\mathrm{mas},  \ee
and taking for simplicity that $\gamma' =  \gamma$, from (\Ref{DT_light}) and  (\Ref{GR_light}) we obtain 
\be
\frac{5\gamma \theta_0^{\rm GR} c^2}{4}\leq \Delta\theta^{exp}.
\ee
and after substitution and solving for $\gamma$, we obtain:
\be
\gamma \leq \frac{4\Delta\theta^{exp}}{5 \theta_0^{\rm GR} c^2}=3.38 \times 10^{-19} \frac{s^2}{m^2},
\ee
where we assume $\theta_0^{\rm GR}$ coincides with the experimental data given in (\ref{theta_exp}), namely 
\be\theta_0^{\rm GR}=\theta^{exp}=31.62 ~ \mathrm{mas}.\ee

From this result, we have estimated the corresponding minimal length for the electron. 
Using Eq.~(\ref{eq:minimal length}) and Eq.~(\ref{beta}), we find:
\begin{equation}
\ell_{\mathrm{min}}^{(e)} = 2 \hbar \sqrt{\beta_e}=\frac{2 \hbar \sqrt{\gamma}}{m_e} \leq 1.35 \times 10^{-13}~\mathrm{m}.
\end{equation} 
This result is naturally weaker than the $ 10^{-16}\,\mathrm{m}$ bound reported in \cite{Stetsko} from Lamb shift measurements in hydrogen atom, reflecting the extremely high precision of hydrogen spectroscopy.

Similarly, for Mercury, the estimated minimal length is
\begin{equation}
\ell_{\mathrm{min}}^{(\mathrm{M})} = 2 \hbar \sqrt{\beta_\mathrm{M}}=\frac{2 \hbar \sqrt{\gamma}}{m_\mathrm{M}}  \leq 3.71 \times 10^{-67}~\mathrm{m}.
\end{equation}
This value differs by about one order of magnitude only from those obtained in \cite{Benczik2002, Samar2020, Kovach2022} based on astrophysical observations of Mercury's orbital precession. 
Interestingly, the minimal length estimates derived from these astrophysical observations, including both Mercury's precession and gravitational lensing, are nearly coincident. 
Although gravitational lensing is less precise, it produces an estimate almost identical to that from the more accurate Mercury precession measurements, suggesting that lensing could serve as a promising method for testing the minimal length hypothesis.

\section{Conclusion}
In this paper, we have explored the Kepler problem in the space of minimal length, which was described through a deformed commutation relation. 

It was shown that taking this minimal length effect into consideration leads to additional corrections to the Hamiltonian, which, in turn, affect the particle's dynamics, specifically, the orbit precession. 

We derived the correction to the scattering angle for unbounded trajectories and demonstrated that the presence of a minimal length scale reduces the scattering angle. An important observation arising from this analysis is the apparent violation of the {weak equivalence principle}, since the scattering angle initially depends on the particle mass. We proposed a resolution to this issue by assuming that the deformation parameters—and hence the minimal length—are mass-dependent. Within this framework, the weak equivalence principle is effectively restored.

This approach also enables a consistent extension of the formalism to massless particles. Applying it to photon scattering in a gravitational field, and using observational data from Einstein rings, we obtained upper bounds on the minimal length for the electron and for Mercury:
\begin{equation}
\ell_{\mathrm{min}}^{(e)} \leq 1.35 \times 10^{-13}~\mathrm{m}, \quad 
\ell_{\mathrm{min}}^{(\mathrm{M})} \leq 3.71 \times 10^{-67}~\mathrm{m}. \nonumber
\end{equation}

Remarkably, the bounds derived from two independent astrophysical phenomena — Mercury’s perihelion precession and gravitational lensing — are in strong agreement. Although gravitational lensing is currently less precise than planetary orbit measurements, it yields nearly identical constraints. This suggests that lensing observations may serve as a robust and independent tool for probing minimal length effects.  
With future improvements in observational precision, the bound obtained from gravitational lensing could be significantly tightened, opening a promising avenue for testing quantum gravity phenomenology through astrophysical observations.

\section*{Acknowledgements}
This work was partially supported by project FF-28F (No. 0126U002265) from the Ministry of Education and Science of Ukraine.

\end{document}